\documentclass{jfm}
\usepackage{graphicx}
\usepackage{epstopdf, epsfig}
\usepackage{bm}
\usepackage{amsmath,amsfonts}
\usepackage{hyperref} 
\usepackage{caption}
\usepackage{subcaption}
\usepackage{csquotes}
\usepackage{xcolor}

\title{Effect of micelle breaking rate and wall slip on unsteady motion past a sphere translating in wormlike micellar solutions}

\author{Chandi Sasmal\aff{1}
  \corresp{\email{csasmal@iitrpr.ac.in}},
  }

\affiliation{\aff{1}Soft Matter Engineering and Microfluidics Lab, Department of Chemical Engineering, Indian Institute of Technology
Ropar, Rupnagar, India-140001}

\begin{document}
\maketitle
\begin{abstract}
In a recent numerical study, we have shown that the unsteady motion past a sphere translating steadily in a wormlike micellar solution is caused due to the breakage of long micelles downstream of the sphere once the Weissenberg number exceeds a critical value based on the two-species Vasquez-Cook-McKinley (VCM) constitutive model for wormlike micelles (C. Sasmal, Unsteady motion past a sphere translating steadily in wormlike micellar solutions: a numerical analysis, Journal of Fluid Mechanics, 912, A52, 2021). This study further shows that this unsteady motion is strongly influenced by the micelle breakage rate and wall slip present on the sphere surface. In particular, we find that the onset of this unsteady motion is delayed to higher values of the Weissenberg number as the micelle breakage rate decreases or, in other words, as the micelles become hard to break. Additionally, we observe that at some values of the micelle breakage rate, a transition in the flow field from steady to unsteady occurs as the Weissenberg number increases, and then again, a transition from unsteady to steady occurs as the Weissenberg number further increases. Therefore, there is a window of the Weissenberg number present in which one can see this unsteady motion past the translating sphere. On the other hand, we show that the presence of wall slip on the sphere surface suppresses this unsteady motion past the translating sphere, and a probable explanation for the same is provided in this study.   
\end{abstract}


\section{\label{Intro}Introduction}
Sedimentation of solid particles in a liquid is ubiquitous in many industrial applications, for instance, hydraulic drilling, fixed and fluidized bed reactors, wastewater treatment, mining and petroleum industries, food processing, etc. Although all of these practical applications deal with a multi-particle system; however, a thorough understanding of the flow characteristics of a single particle is vital to facilitate the corresponding understanding of the complex flow behaviour of these multi-particle systems. Due to this reasoning, the study of a spherical particle falling in a stagnant liquid is considered as one of the benchmark problems in the domain of transport phenomena for many decades. Over the years, a significant amount of studies, comprising of theoretical, numerical, and experimental analysis, have been carried out on this benchmark problem in the literature both for simple Newtonian and complex non-Newtonian fluids. It has been observed that the falling behaviour of a solid particle in a liquid is not only influenced by the confinement of the domain, size and shape of the particles, but also due to the rheological properties of the liquid if it is non-Newtonian in nature like polymer solutions, melts, or wormlike micellar solutions. In the literature, some excellent review articles are present on this complex flow behaviour of a solid particle falling in non-Newtonian fluids~\citep{chhabra2006bubbles,mckinley2002steady,michaelides2006particles}.

One of the complex flow characteristics in the falling behaviour of a spherical particle in viscoelastic polymer solutions is the formation of a negative wake downstream of the sphere. It has been observed both in experiments and numerical simulations once the ratio of the Deborah number to that of the Trouton ratio becomes large~\citep{arigo1998experimental,bush1994stagnation,harlen2002negative,bisgaard1983velocity}. Harlen~\citep{harlen2002negative} proposed that the magnitude of this negative wake downstream of the sphere depends on the relative importance of the extensional flow strength downstream wake of the sphere and the polymer relaxation time due to the shear deformation in the vicinity of the sphere. The ratio of these two phenomena is dictated by the finite extensibility of polymer molecules. Polymers with low finite extensibility allow forming a negative wake, whereas polymers with high extensibility do not. Because of this, a negative wake was observed in shear-thinning polymer solutions, but not in constant viscosity and high elastic Boger fluids~\citep{harlen2002negative,arigo1998experimental}. Along with this negative wake, some studies also found the existence of an unsteady motion of the sphere when falling in viscoelastic polymer solutions. For instance, Bisgaard~\citep{bisgaard1983velocity} observed an unsteady motion of the sphere when sedimenting in a polymer solution comprised of polyacrylamide and glycerol at high Deborah numbers. On the other hand, Binus and Phillips~\citep{binous1999dynamic} also found a similar kind of unsteady motion in their numerical simulations using the FENE (finitely extensible non-linear elastic) spring force law once the Deborah number exceeded a critical value of 5. They showed that this unsteady motion is caused due to the presence of a gradient in dumbbell extensions at the sides of the sphere, leading to the movement of the sphere from a region of highly extended dumbbells to a lower one. Once the sphere reaches this new region with high velocity (due to the less extension of the dumbbells), the dumbbells again get extended, and the sphere velocity decreases. This process repeats with time and causing a periodic motion of the sphere. 

A similar kind of unsteady motion of the sphere and surrounding fluid has also been noticed in many experiments when it falls in a wormlike micellar solution~\citep{chen2004flow,zhang2018unsteady,mohammadigoushki2016sedimentation}. These viscoelastic fluids are formed when surfactants are dissolved in a solvent like water in the presence of a counterion or a salt. Beyond a critical concentration, these surfactants self assemble and form cylindrical aggregate, called micelles. On further increasing the surfactant and salt concentrations, these micelles grow in size, resulting in the formation of wormlike micelles~\citep{moroi1992micelles,dreiss2007wormlike,dreiss2017wormlike}. These worms can entangle and form transient networks, and hence, the solution can exhibit high viscoelastic properties. In a recent experimental study~\citep{wu2021linear}, it has found that even the microstructure of these networks can influence the falling behaviour of a sphere in micellar solution.  The rheological properties of these wormlike micellar (WLM) solutions were found to be more complex than that seen for polymer solutions or melts~\citep{rothstein2008strong,rothstein2003transient}. This is because the transient network formed by these worms can undergo continuous scission and reformation in an imposed flow field. This is in contrast to polymer molecules which are unlikely to break due to the presence of a strong covalent backbone. The reason behind the unsteady motion of a falling sphere in these wormlike micellar solutions is not due to a gradient in the micellar extensions at the sides of the sphere, as proposed by Binus and Phillips~\citep{binous1999dynamic} in the case of the polymer solution. This was experimentally proved by Wu and Mohammadigoushki~\citep{wu2018sphere} based on the flow induced birefringence (FIB) technique. They found a perfect symmetry in the birefringence pattern seen at the sides of the sphere, suggesting that there was no gradient in the micellar extensions at these two regions. This leads to the suggestion that this unsteady motion of the sphere in WLM solutions may be caused due to the sudden rupture of long and extended micelles downstream of the sphere, resulting from an increase in the extensional flow strength downstream of the sphere~\citep{chen2004flow,mohammadigoushki2016sedimentation,wu2018sphere}. However, this was proposed based on some indirect experimental findings seen in the flow-induced birefringence experiments~\citep{wu2018sphere} and the sudden rupture behaviour of WLM solution filaments observed in uniaxial extensional flows carried out in a filament stretching rheometer~\citep{rothstein2003transient}. However, there was no concrete and direct evidence present behind this proposal on this unsteady motion of a sphere falling in WLM solutions. 

Only in a recent study~\citep{sasmal2021}, we showed that this unsteady motion past a sphere is, indeed, caused due to the sudden breakage of long and extended micelles downstream of the sphere based on the numerical simulations using the two-species Vasquez-Cook-McKinley (VCM)~\citep{vasquez2007network} constitutive model for micelles. This model considers the wormlike micelles as an elastic segment
composed of Hookean springs, which all together form an elastic network that can continuously break and reform in a flow field. The breaking and reforming processes of this model were incorporated based on the discrete and simplified version of the Cate’s reversible breaking theory for wormlike micelles~\citep{cates1987reptation}. Our study had considered a problem wherein the sphere is translating at a constant velocity along the axis of a cylindrical tube filled with WLM solutions. This does not represent an exact experimental setting for the falling sphere problem. However, even with this simplified problem, we proved the hypothesis behind the unsteady motion past a sphere in WLM solutions. We also showed a remarkable similarity in the pattern of the unsteady motion generated downstream of the sphere between the present steadily translating sphere and prior experiments performed on the falling sphere. 

To perform the simulations in our earlier study, we imposed the standard no-slip and no-penetration boundary conditions at the sphere surface. However, it is well known that micellar solutions can, very often, exhibit the wall slippage phenomenon during its flow past a solid surface. In this phenomena, the velocity of the wormlike micellar solution does not match with the velocity of the solid wall, and many measurement techniques have been employed to determine this local velocity of WLM solutions like Nuclear Magnetic Resonance (NMR), Particle Tracking Velocimetry (PTV), Particle Image Velocimetry (PIV), etc~\citep{mair1997shear,lopez2006rheo,hu2005kinetics,boukany2008use,mendez2003particle}. The effect of the wall slippage phenomena of WLM solutions is mostly studied on the flow dynamics through a two-dimensional microchannel, circular capillary, or Taylor-Couette flow. It has been observed that the flow physics, particularly the existence of the 'shear banding' phenomena and its stability occurring in these geometries, are greatly influenced by the wall slippage mechanism~\citep{britton1999transition,masselon2008nonlocal,fardin2012shear,fielding2007complex}. However, it should be mentioned here that the flow field in all these geometries is primarily dominated by the shear flow. In the case of translating sphere problem, the flow field is composed of both shear (in the vicinity of the sphere) and extensional flow (downstream of the sphere) dominated regions, and hence it would be interesting to see how the wall slippage mechanism would tend to influence the overall flow field, in particular, the unsteady flow field past the sphere.  As far as the effect of the boundary condition on the phenomena of a falling sphere is concerned, there is only one experimental study available in the literature, performed by Mohammadigoushki and Muller~\citep{HadiBoundaryConditions} with CTAB/NaSal micellar solutions. They found that the wall slip at the roughened sphere surface ( originated due to the presence of air microbubbles on the sphere surface ) leads to a larger terminal velocity than that seen for the smooth sphere. Furthermore, they found that the stagnation point moves closer to the sphere center of mass, and the negative wake shows a stronger magnitude for the roughened sphere than the smooth one. One of our aims of the present study is to numerically investigate the same for the translating sphere problem and validate and explore the findings seen in the corresponding experiments by Mohammadigoushki and Muller~\citep{HadiBoundaryConditions}.

Along with the wall slip, another factor that can influence the flow dynamics around a translating sphere in WLM solutions is the micelle breaking rate and/or how easy or hard it is to break a micelle. This factor can directly modify the rheological properties of a WLM solution like the shear-thinning and extensional hardening phenomena. For instance, if the micelles become easy to break, the shear-thinning properties become dominant, whereas the extensional hardening tendency becomes weaken~\citep{vasquez2007network}. On the other hand, if the micelles become hard to break, a reverse trend can be seen in the rheological properties. These variations in the rheological properties with the micelle breaking rate can govern the flow dynamics of micellar solutions. For instance, Moss and Rothstein~\citep{moss2010flow} investigated the problem of flow of two wormlike micellar solutions, namely, CTAB/NaSal and CPyCl/NaSal, past a microfluidic cylinder placed in a channel, and found the existence of the elastic instability in the former micellar solution but not in the latter one under the same conditions. This was attributed to the fact that CPyCl/NaSal micelles were hard to break as the strain rate downstream of the sphere was not sufficient within the ranges of conditions encompassed in their study. It was also reflected in the response of the micellar solutions in uniaxial extensional flows wherein CPyCl/NaSal solution showed a higher strain hardening than that seen for CTAB/NaSal solution. Likewise, one would expect that micelle breaking rate can also influence the unsteady motion past a sphere as it is also originated due to the breaking of micelles downstream of the sphere~\citep{sasmal2021}, resulting from an increase in the extensional flow strength. Hence, another aim of the present study is to explore the role of this micelle breaking rate on the onset of this unsteady motion past a steadily translating sphere. 

To fulfill the aforementioned objectives, likewise our previous study~\citep{sasmal2021}, we use the two-species VCM constitutive equations to model the rheological behaviour of the present wormlike micellar solutions. The governing equations, namely, continuity, momentum, number density, and conformation tensor evaluations, are solved based the open-source CFD code OpenFOAM for a wide range of the non-dimensional governing parameters.

\section{\label{ProbFor}Problem formulation and governing equations}
\begin{figure}
    \centering
    \includegraphics[trim=4cm 4cm 5cm 1cm,clip,width=12cm]{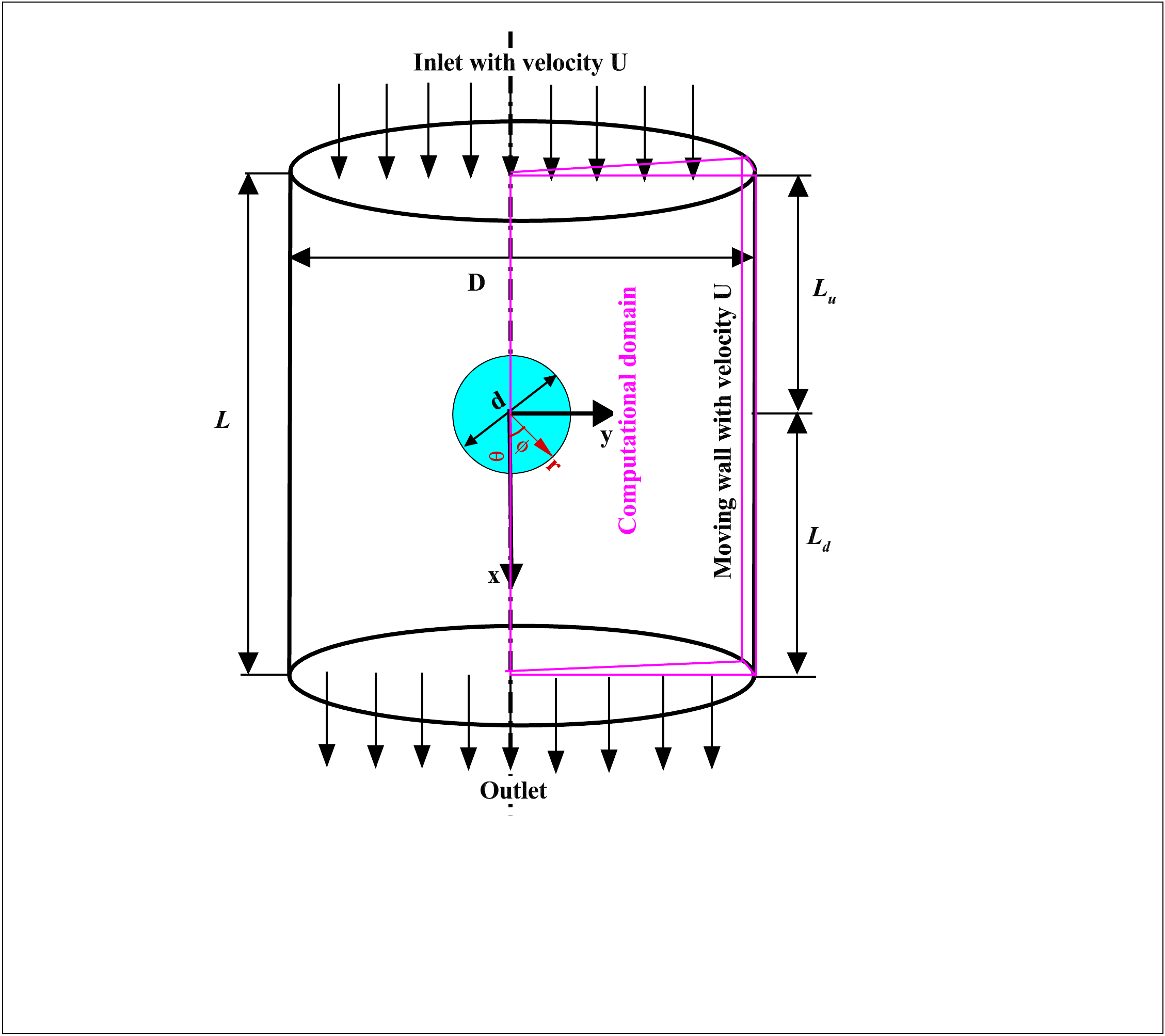}
    \caption{Schematic of the problem considered in this study.}
    \label{fig:figure1}
\end{figure}

The schematic of the problem studied in this study has already been presented in our previous study~\citep{sasmal2021}. However, in the present study, it is again depicted in figure~\ref{fig:figure1}  for the sake of completeness of the problem formulation and to facilitate the discussion of the present new results. As that of our previous study, also in this study, a solid sphere of diameter $d$ is translating steadily along the axis of a cylindrical tube of diameter $D$, which is filled with an incompressible wormlike micellar solution as schematically shown in figure~\ref{fig:figure1}. The study is conducted at a fixed blockage ratio of $BR = \frac{d}{D} = 0.1$. 
 The details of the VCM model can be found out in the study of Vasquez et al.~\citep{vasquez2007network}. In this study, only the salient features of this model are summarized here which are also presented in our earlier study~\citep{sasmal2021}. According to this model, the entangled network of wormlike micellar solution is composed of two elastically active species, namely, a long species $'A'$ and a short species $'B'$. Both these species are modelled as a Hookean elastic segment of length $L$ and $L/2$, respectively. In this model, the long species A continuously breaks and the short species B reforms so that we can have an equation of the form $A \rightleftarrows 2B$. As mentioned in the preceding section, these breakage and reformation dynamics are incorporated based on the discrete version of the Cates 'living polymer' model~\citep{cates1987reptation}. The VCM model provides the non-linear equations for the species conservation equations of the long $(n_{A})$ and short species $(n_{B})$ along with their equations for the evolution of the conformation tensors, namely, $\bm{A}$ and $\bm{B}$. According to this model, the equations for the variation of $n_{A}$, $n_{B}$, $\bm{A}$ and $\bm{B}$ are given in their non-dimensional forms as follows~\citep{vasquez2007network}
\begin{equation}
    \label{nA}
    \mu\frac{Dn_{A}}{Dt} - 2\delta_{A} \nabla^{2}n_{A}  = \frac{1}{2} c_{B} n_{B}^{2} - c_{A}n_{A}
\end{equation}
\begin{equation}
    \label{nB}
    \mu\frac{Dn_{B}}{Dt} - 2\delta_{B} \nabla^{2}n_{B}  = - c_{B} n_{B}^{2} + 2 c_{A}n_{A}
\end{equation}
\begin{equation}
    \label{A}
    \mu \bm{A}_{(1)} + A -n_{A} \bm{I} -\delta_{A} \nabla^{2}\bm{A} = c_{B} n_{B} \bm{B} - c_{A} \bm{A}
\end{equation}
\begin{equation}
    \label{B}
    \epsilon \mu \bm{B}_{(1)} + B -\frac{n_{B}}{2} \bm{I} -\epsilon\delta_{B} \nabla^{2}\bm{B} = -2\epsilon c_{B} n_{B} \bm{B} + 2 \epsilon c_{A} \bm{A}
\end{equation}
Here the subscript $( )_{(1)}$ denotes the upper-convected derivative which is given as $\frac{\partial()}{\partial t} + \bm{U}\cdot \nabla () - \left( (\nabla \bm{U})^{T} \cdot () + ()\cdot \nabla \bm{U}\right)$. The non-dimensional parameters $\mu$, $\epsilon$ and $\delta_{A,B}$ are given as $\frac{\lambda_{A}}{\lambda_{eff}}$, $\frac{\lambda_{B}}{\lambda_{A}}$ and $\frac{\lambda_{A} D_{A,B}}{d^{2}}$ respectively, where $\lambda_{B}$ is the relaxation time of the short species $B$ and $D_{A, B}$ are the dimensional diffusivities of the long and short species $A$ and $B$, respectively. Furthermore, according to the VCM model, the non-dimensional breakage rate $(c_{A})$ of the long species $A$ depends on the local state of the stress field, and it is given by the expression as $c_{A} = c_{Aeq} + \mu \frac{\xi}{3}\left( \dot{\bm{\gamma}}: \frac{\bm{A}}{n_{A}} \right)$, whereas the reforming rate of the long chain species $A$ from the two short species $B$ is assumed to be constant which is given by the equilibrium reforming rate, i.e., $c_{B} = c_{Beq}$. Here the non-linear parameter $\xi$ controls the stress-induced micelle breakage rate. Therefore, it can be seen that the brekage rate of micelles depends on the scalar product of the local stress and shear rate, which ultimately signifying the measure of the local rate of energy dissipation. 

For a non-homogeneous flow field, the aforementioned equations for the evaluation of the micelle number density and conformation tensor must be solved along with the equations for the conservation of mass and momentum, written in their dimensionless form as follows\newline
Mass conservation equation
\begin{equation}
\label{mass}
    \bm{\nabla} \cdot \bm{U} = 0
\end{equation}
Cauchy momentum conservation equation
\begin{equation}
\label{mom}
   El^{-1} \frac{D\bm{U}}{Dt} = -\nabla P + \nabla \cdot \bm{\tau}
\end{equation}
In the above eqs, $\bm{U}$, $t$ and $\bm{\tau}$ are the velocity vector, time and total extra stress tensor respectively, whereas $El$ is the elasticity number defined as the ratio of the Reynolds number to that of the Weissenberg number, i.e., $El = \frac{Wi}{Re}$. Here the Weissenberg number is defined as $Wi = \frac{\lambda_{eff}U}{d}$, whereas the Reynolds number is defined as $Re = \frac{d U \rho}{\eta_{0}}$. In these definations, $\rho$ is the fluid density, $\lambda_{eff}$ is the effective relaxation time, and $\eta_{0}$ is the zero-shear rate viscosity. The total extra stress tensor, $\bm{\tau}$, for a wormlike micellar solution is given as
\begin{equation}
     \bm{\tau} = \bm{\tau}_{w}^{VCM} + \bm{\tau_{s}} = (\bm{A} + 2\bm{B}) - \left(n_{A} + n_{B}\right)\bm{I} + \beta_{VCM}\dot{\bm{\gamma}}
\end{equation}
where $\bm{\tau_{w}}^{VCM}$ is the non-Newtonian contribution from the wormlike micelles, whereas $\bm{\tau_{s}}$ is the contribution from that of the Newtonian solvent which is equal to $\beta \dot{\bm{\gamma}}$. Here the parameter $\beta$ is the ratio of the solvent viscosity to that of the zero-shear rate viscosity of the wormlike micellar solution and $\dot{\bm{\gamma}} = \nabla \bm{U} + \nabla \bm{U}^{T} $ is the strain-rate tensor. Note that here all the lengths, velocity, time and conformation tensors are non-dimensionalized using $d$, $d/\lambda_{eff}$, $\lambda_{eff}$, and $G_{0}^{-1}$ respectively, where $\lambda_{eff} = \frac{\lambda_{A}}{1+c_{Aeq}^{'}\lambda_{A}}$ is the effective relaxation time in the two-species VCM model, $G_{0}$ is the elastic modulus, $\lambda_{A}$ and $c_{Aeq}^{'}$ are the dimensional relaxation time and equilibrium breakage rate of the long species $A$, respectively.

\section{\label{NumDet}Numerical details}
The details of the present numerical setup has already been presented in our earlier study~\citep{sasmal2021} as well as in some other studies~\citep{sasmal2020flow, khan2020effect, khan2021elastic} and hence, it is not again repeated here. Only some salient features are reiterated here for the sake of completeness. While the governing equations, namely, mass, and momentum have been solved using the finite volume method based open-source computational fluid dynamics code OpenFOAM~\citep{weller1998tensorial}, the recently developed rheoFoam solver available in rheoTool~\citep{rheoTool} has been used to solve the VCM constitutive equations. A detailed discussion on the techniques used to discretize different terms of the governing equations has already been presented in our earlier study~\citep{sasmal2021}. The following boundary conditions were employed in order to solve the present problem. On the sphere surface, the standard Navier slip boundary condition was employed, i.e., $\bm{U}_{ws} = -k_{nl} (\bm{n}\cdot \bm{\tau} - [(\bm{n} \cdot \bm{\tau}) \cdot \bm{n}] \bm{n})$ where $k_{nl}$ $(\ge 0)$ is the non-dimensional slip coefficient which controls the extent of the slip velocity on the sphere surface and $\bm{n}$ is the unit normal vector at the solid boundary. The value of $k_{nl}$ depends on the material employed and on the flow conditions, which can be determined based on the fit of the experimental data. Here the subscript 'ws' in the velocity term stands for the 'wall slip'. A no-flux boundary condition is assumed for both the stress and micellar number density, i.e., $\bm{n} \cdot \nabla \bm{A} = \bm{n} \cdot \nabla \bm{B} = 0$ and $\bm{n} \cdot \nabla n_{A} = \bm{n} \cdot \nabla n_{B} = 0$. On the tube wall, $U_{x} = U$ and $U_{y} = 0$, and again no-flux boundary conditions for the stress and micellar number density are imposed. While at the tube outlet, a Neumann type of boundary condition is applied for all variables except for the pressure for which a zero value is assigned here, a uniform velocity of $U_{x} = U$, a zero gradient for the pressure, and a fixed value for the micellar number density are used at the tube inlet. The details of the grid architecture and its number independence study were already presented in our earlier study~\citep{sasmal2021}, and hence those information are not again presented here. 
\begin{figure}
    \centering
    \includegraphics[trim=0cm 0cm 0cm 0cm,clip,width=13cm]{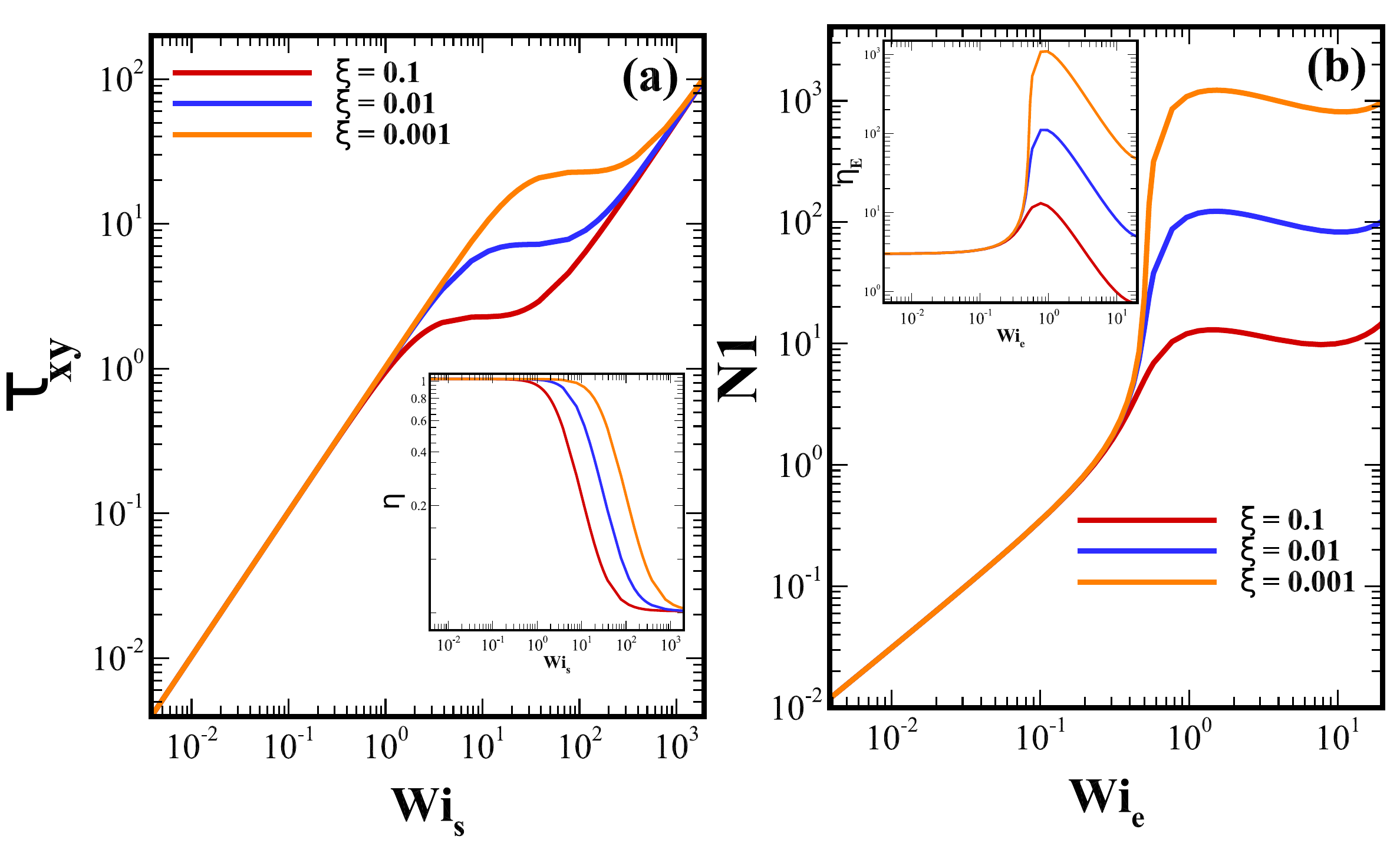}
    \caption{(a) Variation of the non-dimensional shear stress and shear viscosity (inset figure) with the non-dimensional shear rate (shear Weissenberg number, $Wi_{s}$) at different values of the micelle breakage rate (b) Variation of the non-dimensional first normal stress difference and extensional viscosity (inset figure) with the non-dimensional extensional rate (extensional Weissenberg number, $Wi_{e}$) at different values of the micelle breakage rate.}
    \label{fig:rheoflow}
\end{figure}
The rheological responses of the present WLM solution in planar shear and uniaxial extensional flows are presented in figure~\ref{fig:rheoflow} for different values of the micelle breakage rate $\xi$. One can clearly see that the micellar solution shows a shear-thinning property in planar shear flows, whose tendency increases with the increasing value of $\xi$, sub-figure~\ref{fig:rheoflow}(a). On the other hand, the extensional hardening and the subsequent extensional thinning tendencies in uniaxial extensional flows decrease as the micelle breakage rate increases, sub-figure~\ref{fig:rheoflow}(b).

\section{\label{Ressult}Results and discussion}
\subsection{Effect of micelle breakage rate}
\begin{figure}
    \centering
    \includegraphics[trim=5.6cm 7cm 16.2cm 0.5cm,clip,width=12cm]{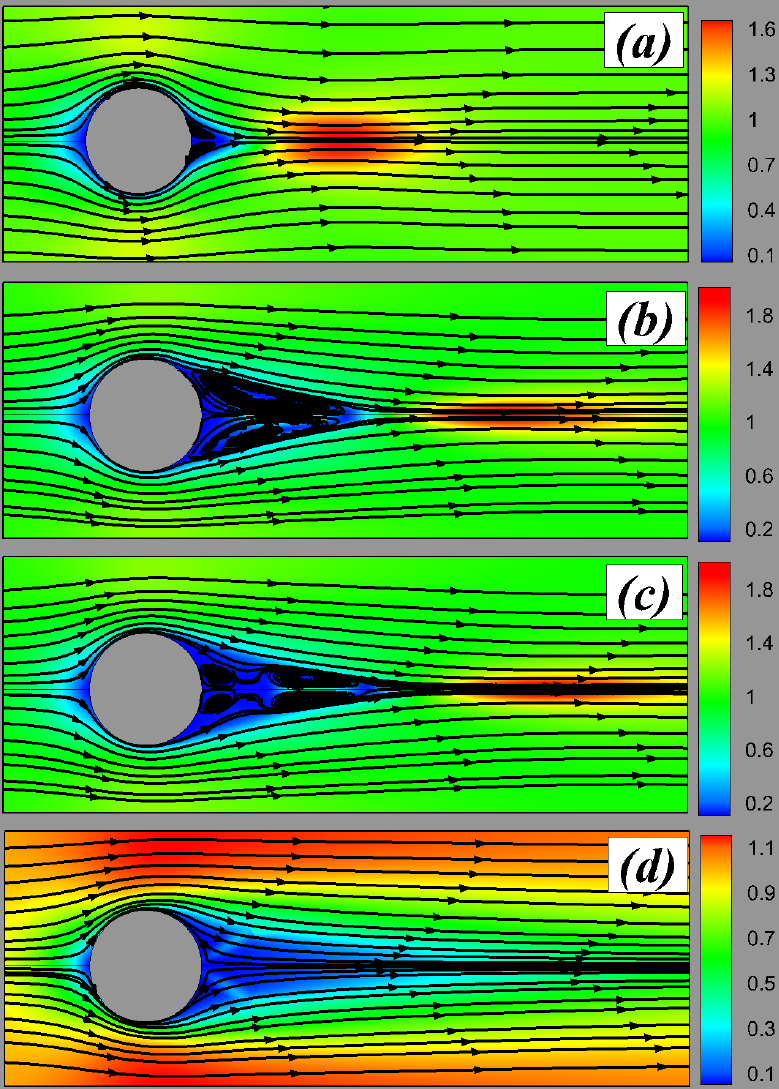}
    \caption{Streamlines and velocity magnitude plots at different values of micelle breakage rate at $Wi = 5$ (a) $\xi = 0.1$ (b) $\xi = 0.01$, $t = 28$ (c) $\xi = 0.01$, $t = 28.1$ (d) $\xi = 0.001$.}
    \label{fig:StreamlinesAtDifferentXi}
\end{figure}
To study the effect of the micelle strength on the flow dynamics around the translating sphere, we use a range of values of the non-linear VCM model parameter $\xi$ $(0.001-0.1)$ in our analysis. The rheological response of the micellar solutions at different values of $\xi$ in planar shear and uniaxial extensional flows are already presented in figure~\ref{fig:rheoflow}. The velocity field in the vicinity of the sphere is visualized in terms of the streamline profiles at different values of $\xi$. Along with it, the surface plot of the velocity magnitude is also presented in the same figure. At low values of the Weissenberg numbers, for instance, at $Wi = 0.1$,  the streamlines are attached to the sphere surface (not presented here), and there is a perfect fore-aft symmetry present both in the streamline and velocity magnitude profiles regardless of the values of $\xi$. This is due to the fact that at this low value of the Weissenberg number, the flow strength is very weak, and hence the breakage and reformation dynamics of the micelles become less important. This results in the flow profiles being the same irrespective of the values of $\xi$. However, as the value of the Weissenberg number and/or the flow strength progressively increases, the flow dynamics become strongly dependent on the micellar breakage rate $\xi$. For instance, a clear difference can be seen at $Wi = 5$ in the streamlines and velocity magnitude plots presented in figure~\ref{fig:StreamlinesAtDifferentXi} at three different values of $\xi$, namely, 0.1, 0.01 and 0.001. At $\xi = 0.1$ (sub-figure~\ref{fig:StreamlinesAtDifferentXi}(a)), one can see that a small vortex is formed just behind downstream of the sphere. 
\begin{figure}
    \centering
    \includegraphics[trim=3cm 6cm 5cm 7.2cm,clip,width=12cm]{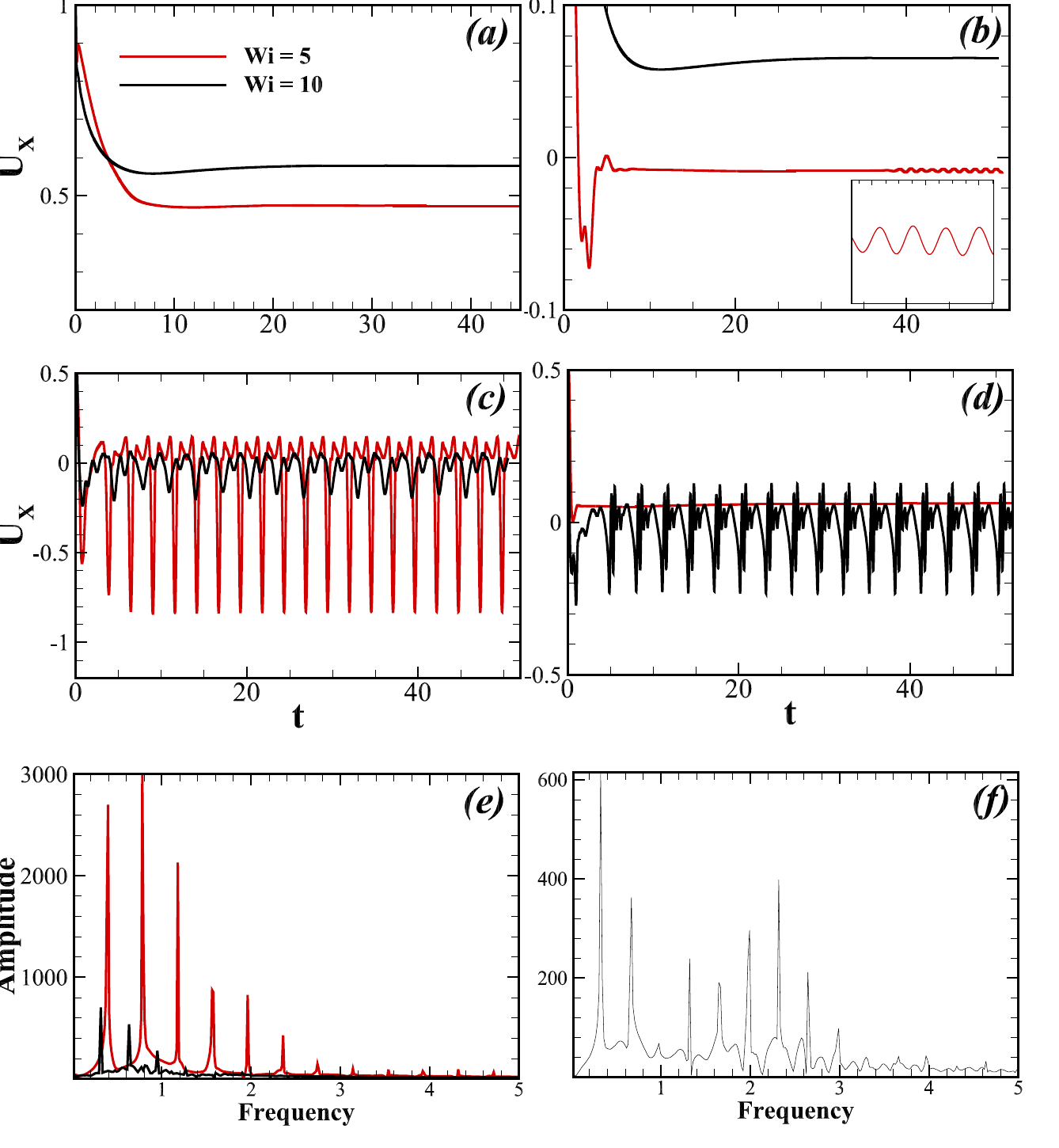}
    \caption{Temporal variation of the non-dimensional axial velocity at a probe location (X = 1.0 and Y = 0) downstream of the sphere at two different Weissenberg numbers, namely, 5 and 10. (a) $\xi = 0.1$ (b) $\xi = 0.04$ (c) $\xi = 0.01$ (d) $\xi = 0.001$. Power spectral density plot of velocity fluctuations at (e) $\xi = 0.01$ and (f) $\xi = 0.001$.}
    \label{fig:VelocityAtDifferentXi}
\end{figure}
\begin{figure}
    \centering
    \includegraphics[trim=0cm 0cm 0cm 0cm,clip,width=12cm]{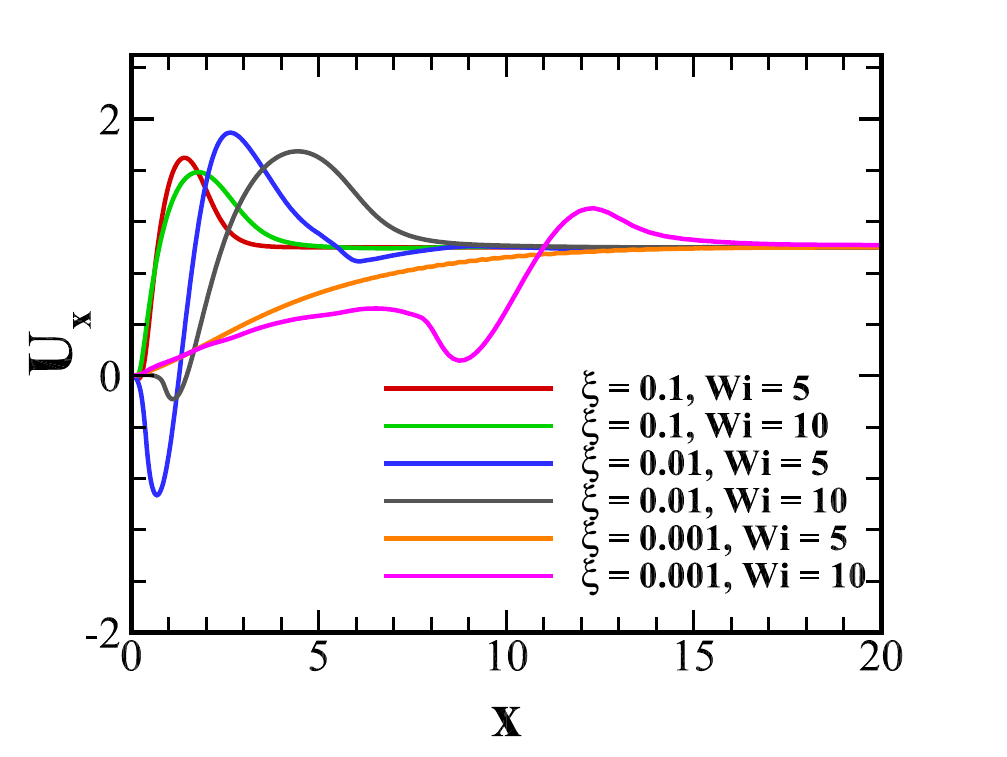}
    \caption{Variation of the non-dimensional axial velocity along the downstream axis of the sphere at different values of micelle breakage rate and two values of the Weissenberg number.}
    \label{fig:VelocityDownstreamAxis}
\end{figure}
Also, a region of high velocity magnitude zone appears downstream of the sphere. This is also evident in figure~\ref{fig:VelocityDownstreamAxis} wherein the non-dimensional stream-wise velocity is plotted along the downstream axis of the sphere. It can be seen that the velocity attains a large value (more than the free stream velocity) at a region almost one sphere diameter away from its rear stagnation point, and then gradually reaches the free stream velocity far away from the sphere. The presence of this 'velocity overshoot' in the velocity profile downstream of the sphere denotes the presence of a 'negative wake' at this region. However, at this value of $\xi$, the velocity field is found to be steady. This is demonstrated in figure~\ref{fig:VelocityAtDifferentXi}(a) wherein the temporal variation of the non-dimensional stream-wise velocity is plotted at a probe location downstream of the sphere ($X = 1.0$ and $Y = 0$), which reaches a steady value with time. At this value of $\xi = 0.1$, the flow even remains steady at a higher Weissenberg number of $Wi = 10$. 

On the other hand, as the value of $\xi$ decreases to 0.01, the flow becomes unsteady at $Wi = 5$. This can be seen in the variation of the streamline profiles presented in sub-figures~\ref{fig:StreamlinesAtDifferentXi}(b) and (c) at two different times. At this value of $\xi$, once again, a recirculation region is formed downstream of the sphere, but its size varies with time. A velocity overshoot is also observed at this value $\xi$ likewise seen at $\xi = 0.1$; however, its magnitude varies with time, for instance, see the results presented at two different times in figure~\ref{fig:StreamlinesAtDifferentXi}. To analyze the nature of the unsteadiness in the flow field past the translating sphere, the temporal variation of the non-dimensional stream-wise velocity is depicted in sub-figure~\ref{fig:VelocityAtDifferentXi}(c) at the same probe location as that presented for $\xi = 0.1$. One can see that the velocity fluctuates with time which is quasi-periodic in nature. This is further confirmed from the power spectrum plot presented in sub-figure~\ref{fig:VelocityAtDifferentXi}(e) wherein the velocity fluctuations are governed by more than one dominant frequencies. A transition in the velocity field from steady to unsteady periodic and then to unsteady quasi-periodic is observed as the Weissenberg number gradually increases, as seen in our earlier study~\citep{sasmal2021} as well as in prior experiments for the falling sphere problem~\citep{zhang2018unsteady,mohammadigoushki2016sedimentation}. A velocity overshoot is, once again, seen along the downstream axis of the sphere at this value of $\xi$, but its magnitude varies with time due to the presence of the unsteadiness in the flow field, figure~\ref{fig:VelocityAtDifferentXi}. As the Weissenberg number further increases to 10, the magnitude of the velocity overshoot decreases along the downstream axis of the sphere, thereby showing a decrease in the tendency of appearing the negative wake downstream of the sphere. The magnitude of the velocity fluctuations at the same probe location as that presented at $Wi = 5$ also decreases, as can be seen from sub-figure~\ref{fig:VelocityAtDifferentXi}(c). This is also evident in the frequency spectrum of the velocity fluctuations presented in sub-figure~\ref{fig:VelocityAtDifferentXi}(e). At $\xi = 0.04$, the velocity field even becomes completely steady at $Wi = 10$, whereas it is fluctuating at $Wi = 5$ although the magnitude of fluctuations is less as compared to that seen at $\xi = 0.01$, sub-figure~\ref{fig:VelocityAtDifferentXi}(b). As the micelle breakage rate further decreases to 0.001, the velocity field is seen to be steady at $Wi = 5$, whereas it becomes quasi-periodic at $Wi = 10$, as can be evident from the plots of non-dimensional stream-wise velocity at the same probe location as that presented for $\xi = 0.01$ (sub-figure~\ref{fig:VelocityAtDifferentXi}(d)) and power spectrum plot of the velocity fluctuations shown in sub-figure~\ref{fig:VelocityAtDifferentXi}(f). 
\begin{figure}
    \centering
    \includegraphics[trim=0cm 0cm 0cm 0cm,clip,width=12cm]{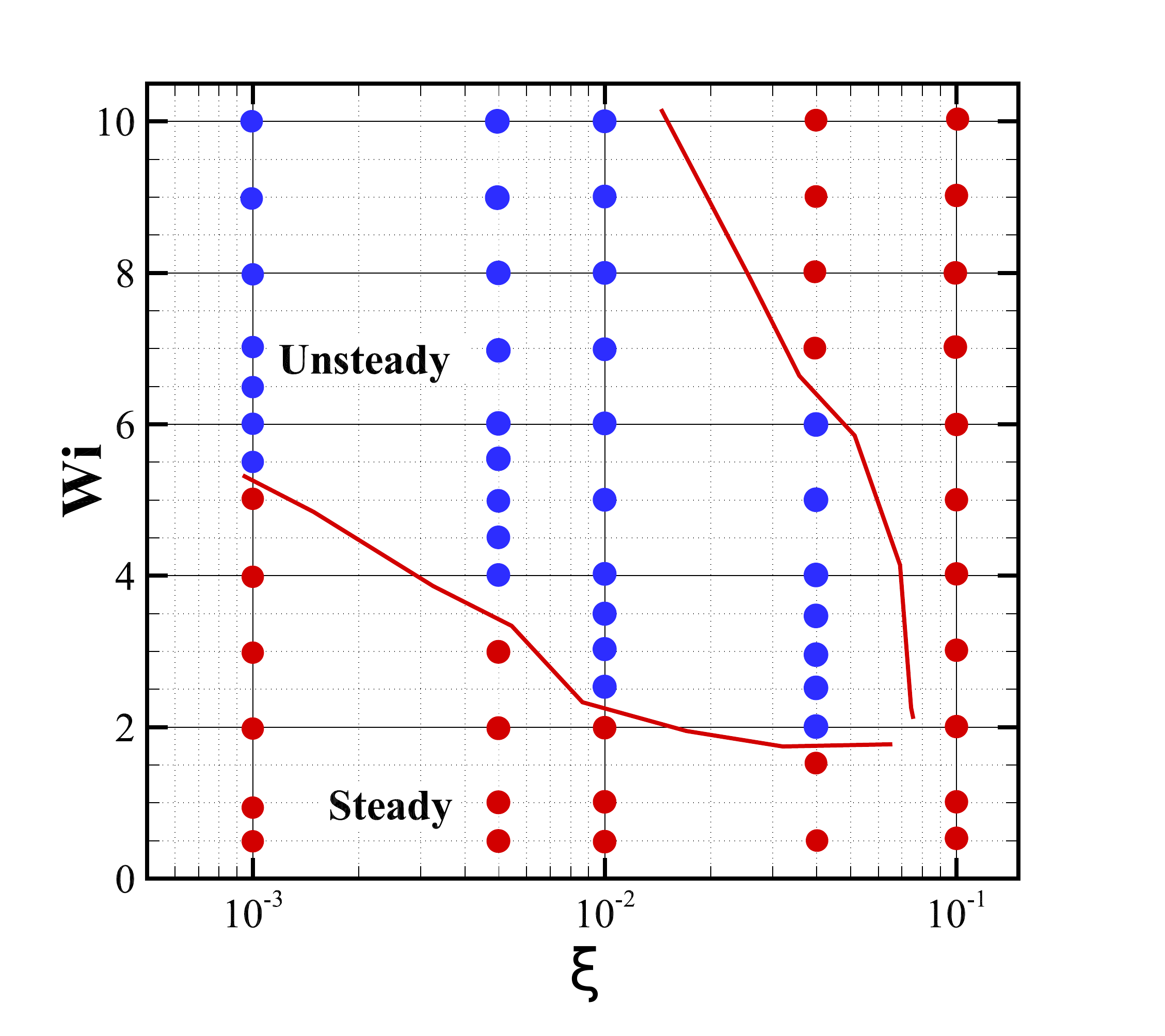}
    \caption{Phase diagram of different flow states with the micelle breakage rate and Weissenberg number.}
    \label{fig:PhaseDiagram}
\end{figure}
A complete phase diagram on the existence of different flow states at different Weissenberg numbers and micelle breakage rates is depicted in figure~\ref{fig:PhaseDiagram}. From this phase diagram and from the above discussion, it can be clearly seen that the onset of unsteady motion past the translating sphere is delayed to higher values of the Weissenberg number as the micelle breakage rate decreases or in other words, as the micelles become progressively harder to break. This is expected as one requires higher flow strength to break a micelle once the micelle breakage rate decreases. It is also clear from figure~\ref{fig:MicelleNumberDensity} wherein the surface plot of long micelle number density in the vicinity of the sphere as well as its variation along the downstream axis are presented for two values of $\xi$, namely, 0.01 and 0.001 at the same Weissenberg number of 3. From both these sub-figures, one can see that at $\xi = 0.01$, more long micelles are broken into smaller ones than that occurred at $\xi = 0.001$
\begin{figure}
    \centering
    \includegraphics[trim=0cm 0cm 0cm 0cm,clip,width=12cm]{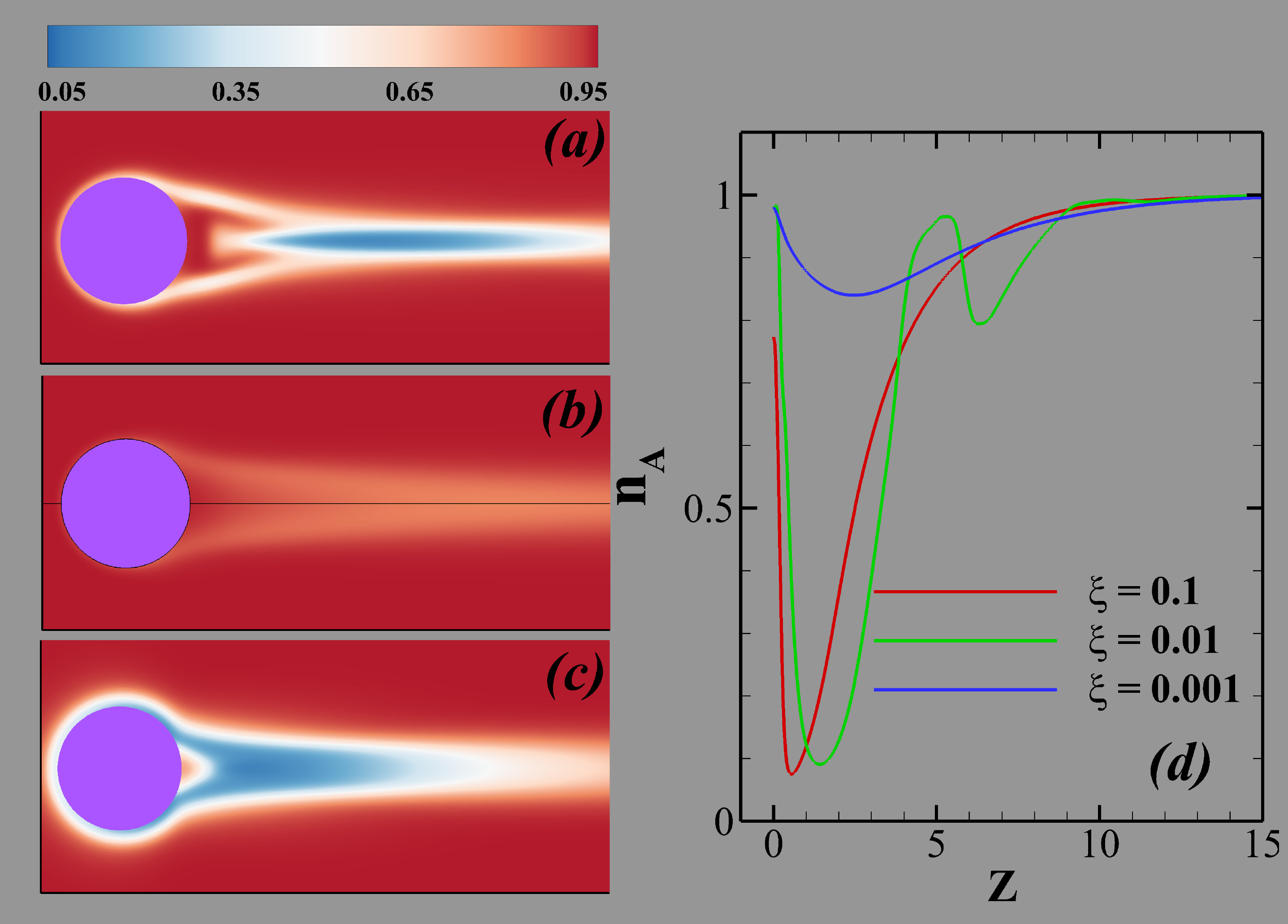}
    \caption{Surface plot of number density of long micelles for different micelle breakage rates at $Wi = 5$. (a) $\xi = 0.01$ (b) $\xi = 0.001$ (c) $\xi = 0.1$. (d) Variation of number density of long micelles along the downstream axis for different micelle breakage rates.}
    \label{fig:MicelleNumberDensity}
\end{figure}
\begin{figure}
    \centering
    \includegraphics[trim=0cm 0cm 0cm 0cm,clip,width=12cm]{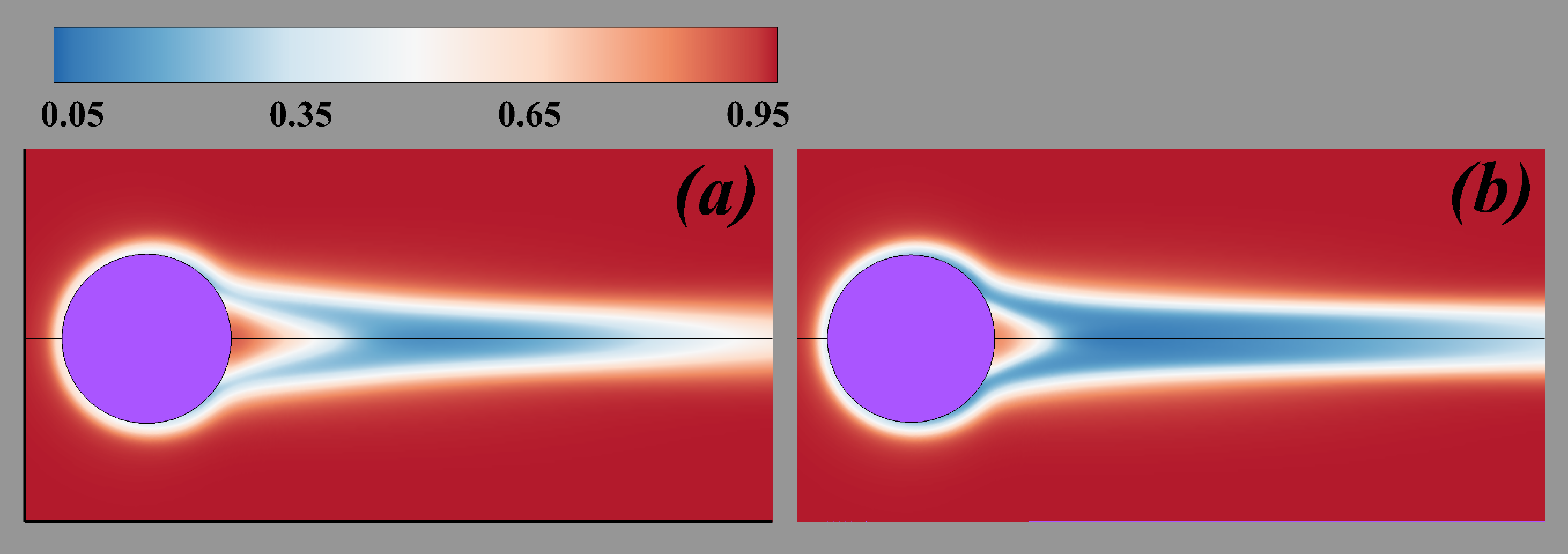}
    \caption{Surface plot of the number density of long micelles for micelle breakage of $\xi = 0.04$ at two different Weissenberg numbers, namely, (a) Wi = 5 and (b) Wi = 10.}
    \label{fig:MicelleNumberDensity_xi0p04}
\end{figure}
under otherwise identical conditions. This results in the flow is to be unsteady at $\xi = 0.01$, whereas it is steady at $\xi = 0.001$. The mechanism behind the origin of this unsteady motion has already been explained in detail in our earlier study~\citep{sasmal2021}. These results, once again, further confirm the hypothesis that the unsteady motion past a translating sphere in WLM solutions is, indeed, originated due to the breakage of stretched and long micelles downstream of the sphere. As the micelle breakage rate increases to $\xi = 0.1$, they become easy to break. Therefore, most of the micelles are broken in the vicinity of the sphere before reaching its downstream section due to the presence of a strong shearing flow field in this region, see sub-figure~\ref{fig:MicelleNumberDensity}(c). This makes the flow to be a steady one at this value of $\xi$. This breakage of long micelles in the vicinity of the sphere also happens at higher Weissenberg numbers, for instance, at $Wi = 10$ (see sub-figure~\ref{fig:MicelleNumberDensity_xi0p04}(b)), and that is why the flow again transits to a steady one, as observed at $\xi = 0.04$. This upper critical value of the Weissenberg number at which the flow becomes again periodic and finally steady increases as the micelles become stiff to break. Although this transition from a quasi-periodic to periodic and finally steady state at higher Weissenberg numbers was not observed in earlier experiments for the falling sphere problem; however, a similar kind of observation was seen in the case of flow past a microcylinder placed in a microchannel. For the flow of a CPyCl/NaSal WLM solution past this geometry, Haward et al.~\citep{haward2019flow} found that as the Weissenberg number gradually increases, the flow transits from a quasi-periodic state to periodic like behaviour after a critical value of the Weissenberg number instead of observing a more chaotic or turbulent like behaviour at higher Weissenberg numbers. They suggested a similar hypothesis for this, which is due to the breakage of long micelles around the microcylinder.           

\begin{figure}
    \centering
    \includegraphics[trim=0cm 0cm 0cm 0cm,clip,width=12cm]{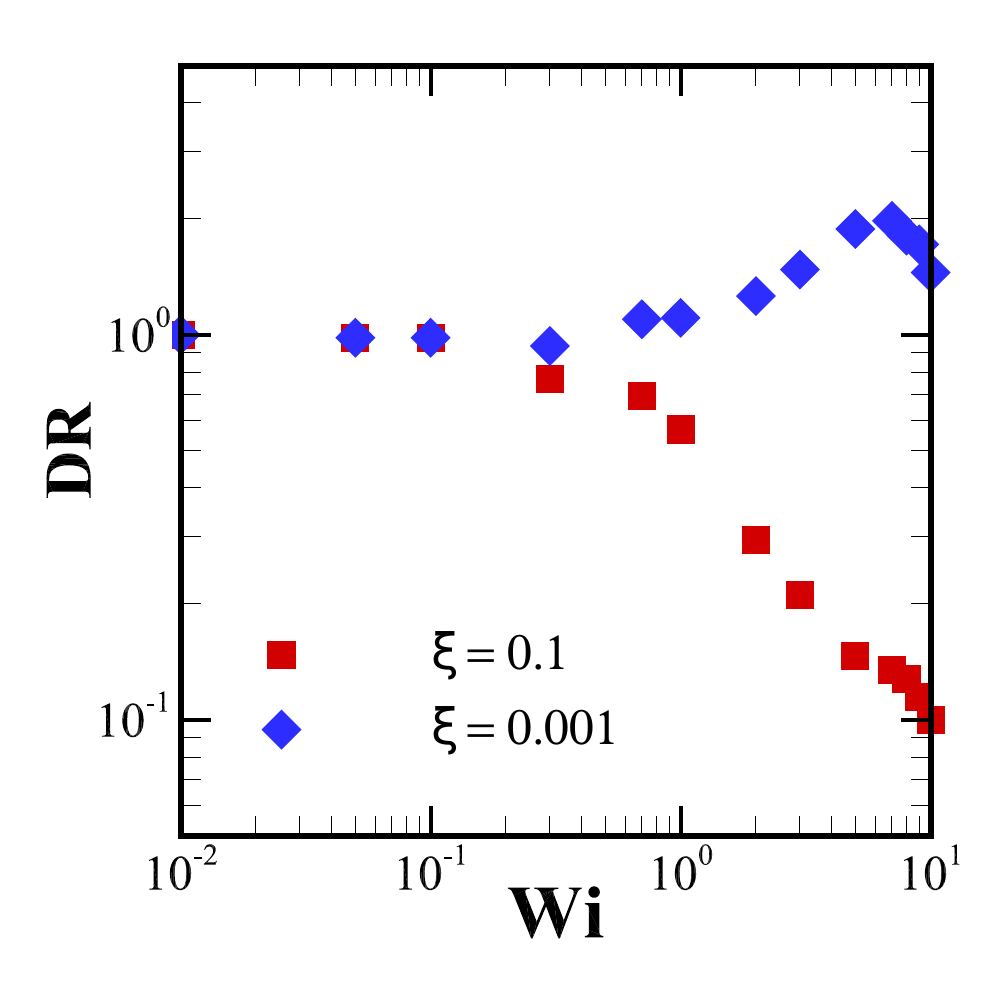}
    \caption{Variation of the drag ratio (DR) with the Weissenberg number and micelle breakage rate.}
    \label{fig:drag}
\end{figure}

Next, the drag forces acting on the translating sphere are presented in terms of the variation of the drag ratio $DR$ (ratio of the drag forces in wormlike micellar solution to that in a Newtonian fluid under otherwise identical condition) with the Weissenberg number at different micelle breakage rates in figure~\ref{fig:drag}. At low values of the Weissenberg number, the drag ratio value shows a value of nearly equal to unity regardless of the values of the micelle breakage rate, and it remains at this value (or shows a plateau) up to a value of the Weissenberg number of about 0.1. This is due to the fact that at this condition, the wormlike micellar solution behaves like a Newtonian fluid, and also, the micelle breakage rate hardly matters due to the existence of low flow strength. However, as the Weissenberg number further gradually increases, the difference becomes significant depending upon the value of the micelle breakage rate. At high values of the micelle breakage rate, for instance, at $\xi = 0.1$, one can see that the drag ratio gradually decreases after showing the plateau region at low Weissenberg numbers. This is mainly due to the dominance of the shear-thinning properties of the micellar solution at this value of $\xi$. Although the same trend is seen for $\xi = 0.01$; however, the values of the drag ratio increase with the decreasing value of $\xi$. This is due to the increase in the contribution of the elastic forces to the drag forces as the micelles become hard to break. On the other hand, at $\xi = 0.001$, the trend in the variation of the drag ratio with the Weissenberg number is somewhat different and complex than that seen for other values of $\xi$. 
At this value of $\xi$, the drag ratio shows a plateau, likewise other values of $\xi$ at low values of the Weissenberg number. It then slightly decreases and then increases and shows a local maximum and finally again decreases as the Weissenberg number progressively increases. A similar kind of trend in the drag ratio has also been observed by Chen and Rothstein~\citep{chen2004flow} in their experiments for the falling sphere problem. The reason behind this trend can be, at least, qualitatively explained as follows: there are two competitive effects present which influence the drag ratio. One is the shear-thinning effect of the micellar solution, which tends to decrease the drag, and another is the elastic stress generated downstream of the sphere, which tends to increase the drag. Initially, at low Weissenberg numbers, the former effect is dominant than the latter one, and hence the drag ratio decreases. However, with the further increment in the Weissenberg number, the elastic stresses become significant, which eventually overcome the shear-thinning effect and increase the drag ratio, thereby showing a maximum in the trend. As the Weissenberg number is further increased, the breakage of long micelles into short ones occurs, thereby reducing the elastic stresses generated downstream of the sphere, which in turn decreases the drag forces acting on the sphere.

\subsection{Effect of wall slip}
\begin{figure}
    \centering
    \includegraphics[trim=0cm 0cm 0cm 0cm,clip,width=13cm]{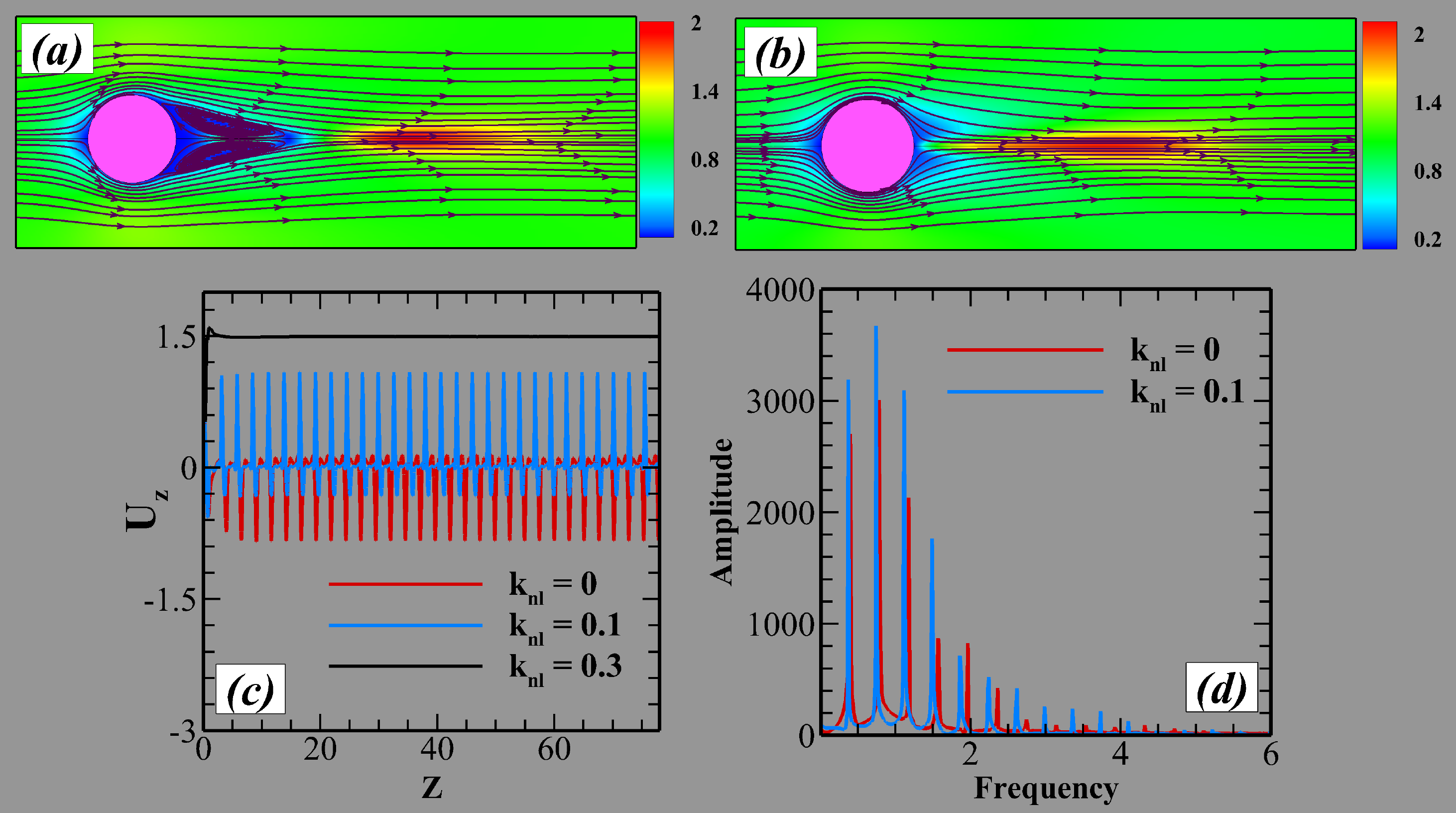}
    \caption{Streamlines and velocity magnitude plots at $Wi = 5.0$ for (a) No-slip, $k_{nl} = 0$ and (b) Slip, $K_{nl} = 0.3$ boundary conditions. (c) Temporal variation of the stream-wise non-dimensional velocity at a probe location (X = 1.0, Y = 0) downstream of the sphere and (d) power spectrum plot of velocity fluctuations for different slip coefficients.}
    \label{fig:VelSlipNoSlip}
\end{figure}

\begin{figure}
    \centering
    \includegraphics[trim=0cm 0cm 0cm 0cm,clip,width=13cm]{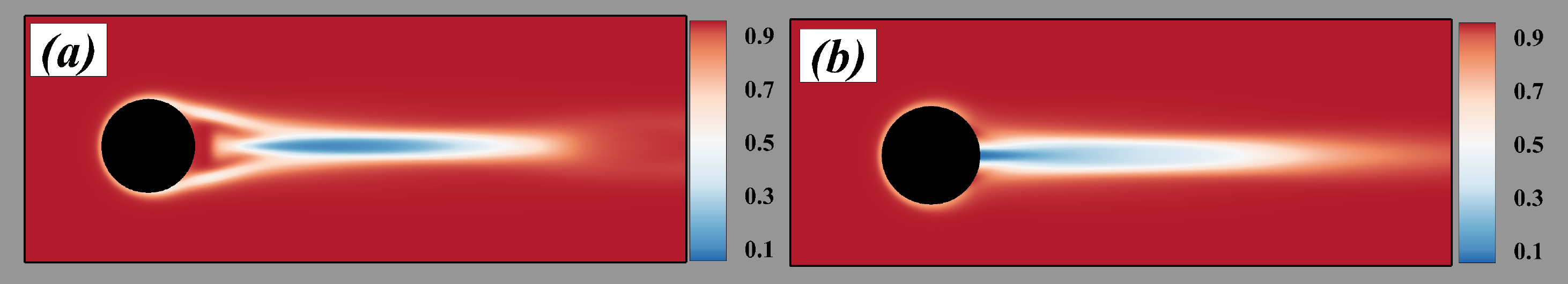}
    \caption{Surface plot of number density of long micelles for two different slip coefficients, namely, 0 (a) and 0.3 (b) at $Wi = 5.0$ and $\xi = 0.01$}
    \label{fig:MicelleNumberDensitySlipNoSlip}
\end{figure}

\begin{figure}
    \centering
    \includegraphics[trim=0cm 0cm 0cm 0cm,clip,width=13cm]{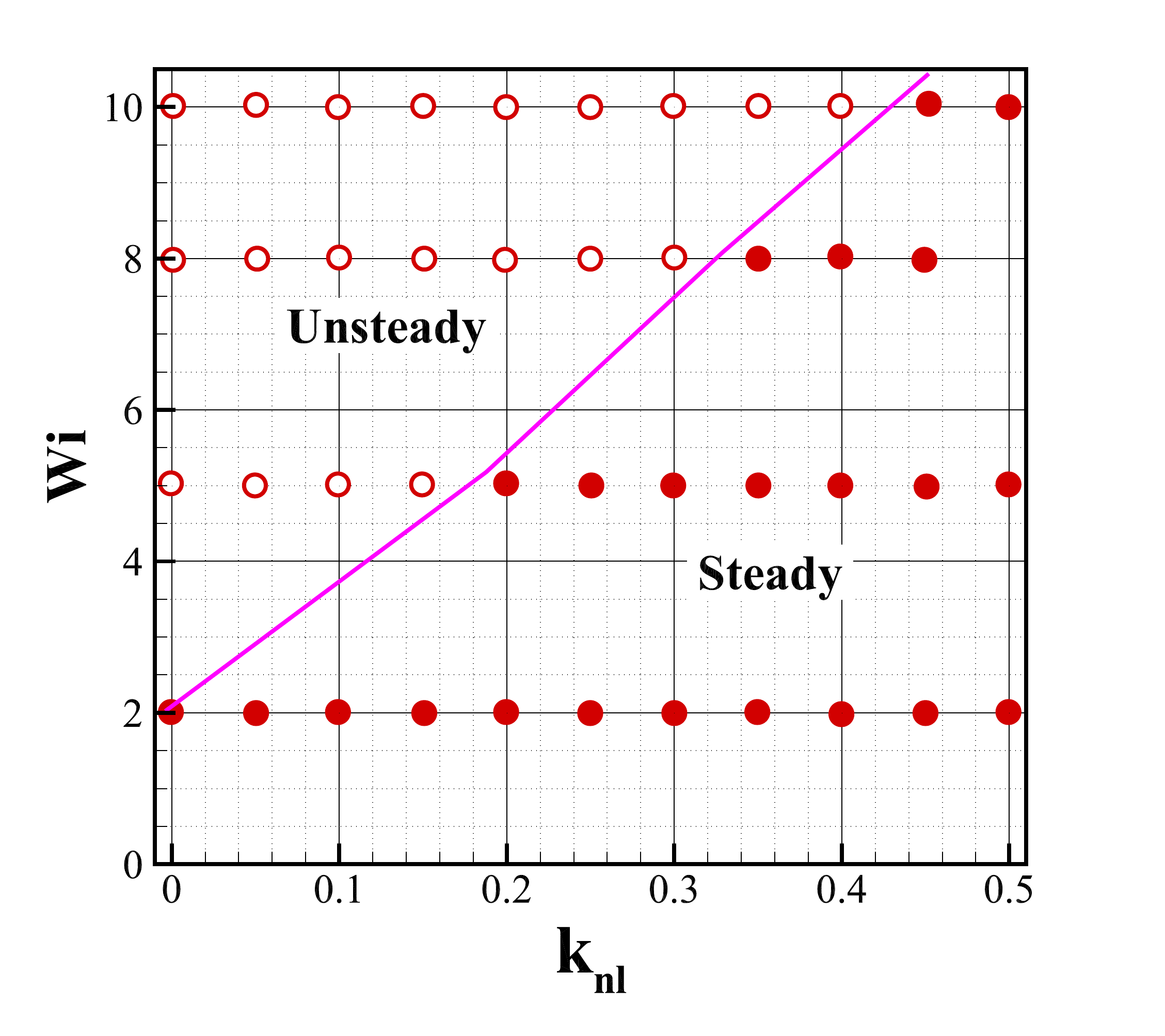}
    \caption{Phase diagram on the existence of unsteady and steady flow fields as a function of the Weissenberg numbers and slip coefficients.}
    \label{fig:PhaseDiagramSlip}
\end{figure}

The effect of wall slip on the flow characteristics of the translating sphere is discussed at a fixed value of the micelle breakage rate of $\xi = 0.01$. The streamline and velocity magnitude plots for two values of the slip coefficients, namely, $k_{nl} = 0$ (no-slip) and 0.3 at $Wi = 5.0$ are presented in sub-figures~\ref{fig:VelSlipNoSlip}(a) and (b), respectively. One can clearly see that the imposition of slip boundary condition on the sphere surface suppresses the separation of the shear boundary layer, thereby showing no recirculation region downstream of the sphere,  as opposed to that seen for the no-slip boundary condition. The corresponding temporal variation of the stream-wise velocity component at a probe location downstream of the sphere is shown in sub-figure~\ref{fig:VelSlipNoSlip}(c) for three values of the slip coefficients, namely, 0, 0.1, and 0.3. At the highest value of the slip coefficient, i.e., at $k_{nl} = 0.3$, it can be seen that the velocity reaches a constant value with time, thereby suggesting that the flow field becomes steady. On the other hand, for the other two values of the slip coefficients, i.e., at 0 and 0.1, the flow field shows a quasi-periodic nature. This can also be seen from the power spectrum of velocity fluctuations presented in sub-figure~\ref{fig:VelSlipNoSlip}(d), wherein more than one dominant frequencies are present at these two values of the slip coefficient. Also, it can be observed that as the slip coefficient increases, the velocity magnitude increases. This is inline with that seen in the literature~\citep{HadiBoundaryConditions}. 
Therefore, it is seen that as the magnitude of the slip coefficient increases, the tendency of appearing unsteady motion downstream of the translating sphere decreases, or even it can be completely suppressed. The reason behind this can be explained as follows: as the slip coefficient on the sphere surface gradually increases, the magnitude of the velocity gradient in the vicinity of the sphere surface decreases. As a result, the long micelles can easily slip over the sphere surface and travel from the front stagnation point towards the rear without much stretching and breaking into shorter micelles. This is also evident in figure~\ref{fig:MicelleNumberDensitySlipNoSlip} wherein the number density of long micelles for two values of the slip coefficient (0 and 0.3) is depicted at $Wi = 5$. It is clear from this figure that the number density of long micelles is more at $k_{nl} = 0.3$ than that seen at $k_{nl} = 0$ in the vicinity of the sphere surface. 
However, once the long micelles reach the rear stagnation point of the sphere, they break into short micelles, thereby forming a strand of short micelles just downstream of the sphere close to the rear stagnation point, as can be seen from sub-figure~\ref{fig:MicelleNumberDensitySlipNoSlip}(b). The formation of this strand of short micelles downstream of the sphere is due to the presence of a strong extensional flow field in this region. Therefore, the suppression of vortex formation and the breakage of long micelles just near the rear stagnation point of the sphere inhibit the onset of unsteady motion downstream of the sphere at finite values of the slip coefficient. This is opposed to that seen for the no-slip condition wherein the breakage of long micelles occurs away from the rear stagnation point of the sphere due to the presence of vortex just behind the sphere~\citep{sasmal2021}. Also, for the no-slip condition, the stretching of long micelles occurs in the vicinity of the sphere due to the presence of a strong velocity gradient in this region. A complete phase diagram on the existence of steady and unsteady flow fields for different values of the Weissenberg number and slip coefficients is shown in figure~\ref{fig:PhaseDiagramSlip}. One can clearly see from this figure that the Weissenberg number, at which the transition from steady to unsteady flow field occurs, increases as the slip coefficient increases.   

\section{\label{Con}Conclusions}
This study is an extension of our earlier study in which we showed that the unsteady motion past a solid sphere translating steadily in wormlike micellar solutions was caused due to the breakage of long micelles downstream of the sphere, resulting from an increase in the extensional flow strength in that region once the Weissenberg exceeded a critical value (C. Sasmal, Unsteady motion past a sphere translating steadily in wormlike micellar solutions: a numerical analysis, Journal of Fluid Mechanics, 912, A52, 2021). A two-species Vasquez-Cook-McKinley (VCM) constitutive model was used to realize the rheological behaviour of the micellar solutions. In this study, we have further shown that this unsteady motion past the translating sphere could be greatly influenced by the micelle breakage rate using the same two-species VCM model. In particular, it has been found that the onset of this unsteady motion is delayed to higher values of the Weissenberg number as the micelles become hard to break and/or its breakage rate decreases. On the other hand, if the breakage rate increases and/or the micelles become very easy to break, one may even not find any unsteady motion at all. Likewise our earlier study, we have again found a gradual transition in the flow field from steady to unsteady as the Weissenberg number progressively increases. However, we have found that as the Weissenberg number is further incremented to very high values, the velocity field transits from unsteady to steady one instead of showing a more chaotic and irregular flow field for certain values of the micelle breakage rate. Therefore, there is a window of the Weissenberg number present in which one can observe the presence of this unsteady motion past the translating sphere. Although we have found a great qualitative similarity in the flow characteristics between the present results for a translating sphere and prior many experimental results for a falling sphere problem, this transition from unsteady to steady flow field at high values of the Weissenberg number was not observed in the experiments carried out for the latter problem. However, a similar kind of transition was observed in the experiments on the flow of wormlike micellar solutions past a microfluidic cylinder confined in a channel at very high values of the Weissenberg number~\citep{haward2019flow}. In this study, we have provided a possible explanation for this behaviour of the micellar solution based on the breakage and reformation dynamics of the micelles in the vicinity and downstream of the translating sphere. 

On the other hand, we have found that the presence of wall slip on the sphere surface increases the velocity downstream of the sphere, likewise it was observed in the experiments for the falling sphere problem~\citep{HadiBoundaryConditions}. Most importantly, it has been observed that the wall slip suppresses the unsteady motion downstream of the sphere. This is possibly due to the slippage of the micelles on the sphere surface and the suppression of the vortex formation downstream of the sphere. Therefore, it would be interesting to carry out some corresponding experiments with different slip lengths on the sphere surface and investigate how it would tend to influence the unsteady motion of both the sphere and the surrounding flow field.     

\section{Acknowledgements}
Author would like to thank IIT Ropar for providing funding through the ISIRD research grant (Establishment1/2018/IITRPR/921) to carry out this work. 

\section{Declaration of interests}
The authors report no conflict of interest

\bibliographystyle{jfm}
\bibliography{jfm-instructions}

\end{document}